 \documentclass[preprint ,showpacs,prd,amsfonts,amsmath,amssymb,floatfix]{revtex4}

\usepackage{hhline}
\usepackage{mathrsfs}
\usepackage{graphicx}
\usepackage{dcolumn}
\usepackage{bm}

\input{epsf}
\def\VEC#1{\mbox{\boldmath $#1$}}
\def\dbar#1{\bar{\bar #1}}

\def\SE{S_\mathrm{e}}
\def\SM{S_\mathrm{m}}
\def\ME{M_\mathrm{e}}
\def\MM{M_\mathrm{m}}
\def\AE{a_\mathrm{e}}
\def\BE{b_\mathrm{e}}
\def\CE{c_\mathrm{e}}
\def\XE{x_\mathrm{e}}
\def\YE{y_\mathrm{e}}
\def\ZE{z_\mathrm{e}}
\def\AM{a_\mathrm{m}}
\def\BM{b_\mathrm{m}}
\def\CM{c_\mathrm{m}}
\def\XM{x_\mathrm{m}}
\def\YM{y_\mathrm{m}}
\def\ZM{z_\mathrm{m}}
\def\TAE{\widetilde{\AE}}
\def\TBE{\widetilde{\BE}}
\def\TCE{\widetilde{\CE}}
\def\TXE{\widetilde{\XE}}
\def\TYE{\widetilde{\YE}}
\def\TZE{\widetilde{\ZE}}
\def\TAM{\widetilde{\AM}}
\def\TBM{\widetilde{\BM}}
\def\TCM{\widetilde{\CM}}
\def\TXM{\widetilde{\XM}}
\def\TYM{\widetilde{\YM}}
\def\TZM{\widetilde{\ZM}}
\def\RE{\VEC{r}_\mathrm{e}}
\def\RM{\VEC{r}_\mathrm{m}}
\def\TE{\VEC{t}_\mathrm{e}}
\def\TM{\VEC{t}_\mathrm{m}}
\def\EE{\VEC{e}_\mathrm{e}}
\def\EM{\VEC{e}_\mathrm{m}}
\def\UE{\VEC{u}_\mathrm{e}}
\def\UM{\VEC{u}_\mathrm{m}}
\def\TRE{\widetilde{\RE}}
\def\TRM{\widetilde{\RM}}
\def\TTE{\widetilde{\TE}}
\def\TTM{\widetilde{\TM}}
\def\TEE{\widetilde{\EE}}
\def\TEM{\widetilde{\EM}}
\def\TUE{\widetilde{\UE}}
\def\TUM{\widetilde{\UM}}
\def\EEX{{\EE}_{ \bar{x}}}

\def\EMX{{\EM}_{\dbar{x}}}

\def\DEE{\dot{\EE}}
\def\DEM{\dot{\EM}}
\def\DEEX{\DEE_{ \bar{x}}}
\def\DEEY{\DEE_{ \bar{y}}}
\def\DEEZ{\DEE_{ \bar{z}}}
\def\DEMX{\DEM_{\dbar{x}}}
\def\DEMY{\DEM_{\dbar{y}}}
\def\DEMZ{\DEM_{ \bar{z}}}

\begin{document}

\title{
Twelve-dimensions theory in electromagnetic dynamics as a hypothesis
}

\author{Yoshiro Nohara}

\date{\today}

\begin{abstract}
We propose a theory in electromagnetic dynamics, 
  in which time and space are equivalent with each other and have totally twelve dimensions.
Then, we solve that with realistic assumptions and find a steady state as a solution.
The solution draws a circle in complex space having six dimensions as a closed string,
  which is an orbit passed by waves of two scalar fields having electric and magnetic characters, respectively.
We also discuss a possibility of existing solutions of electromagnetic waves.
\end{abstract} 

\maketitle 
\section{INTRODUCTION}
Recently, as unified field theories, 
  super string, M, and F theories are proposed and are powerfully studied. \cite{UFT,FTheory}
Those unified field theories are hoped to clear the mysteries of the particle physics,
  in which it begins to be discussed with extra dimensions. \cite{PT}
The unified field theories need additional dimensions except four dimensions of space and time.
Although those extra dimensions are not well known yet, those are expected to be clear experimentally.
There are also some proposals that extra dimensions have a possibility to include time-like dimensions. \cite{ET1,ET2,ET3,ET4}
Actually, in that assumption, some problems in theoretical physics disappear, but it also brings us a possibility of violating causality.
On the other hand, in electromagnetic dynamics, 
  complex space, time, and fields are also proposed,
  and which are not only useful to avoid numerical difficulties but also helpful to predict unknown physics. \cite{CST1,CST2}
Moreover, a possibility of the existence of scalar fields caused by broken Lorentz condition
  begins to be discussed in both the electromagnetic dynamics and quantum gravity. \cite{SF1,SF2,SF3,SF4,SF5,SF6,QG}

As above listed proposals, there are many possibilities of existences of incredible phenomena.
The experiments, to know if those are correct or not, begin to be done in recently.
Therefore, it is important to know what will be happen and to prepare many those predictions.
So, we propose a theory having a potential to help discussing about those phenomena.

\section{DERIVATION}
\subsection{Basic equations}
We first prepare two scalar fields (V/m),
\begin{eqnarray}
 \xi &\mathrm{and}& \eta,
\end{eqnarray}
  which have electric and magnetic properties, respectively.
So, we call those ``electroscalar" and ``magnetoscalar" fields.
Here, we note that, for the convenience, the unit of time $t$ (s) is changed to (m) by multiplying light speed $c$ (m/s) as following formula,
\begin{eqnarray}
t^\mathrm{new} \equiv c t.
\end{eqnarray}
Then, we consider those scalar fields' waves and local coordinates which move along those waves with constant speeds.
Hereafter, we call those waves and systems having the local coordinates the basic waves and local systems, respectively.
We define those local time (m), 
\begin{eqnarray}
s_\mathrm{e} &\mathrm{and}& s_\mathrm{m}, 
\end{eqnarray}
  at certain positions in the local systems.
At the positions, the basic waves are described by
\begin{eqnarray}
 \xi(s_\mathrm{e}) &\mathrm{and}& \eta(s_\mathrm{m}).\label{eq_BW}
\end{eqnarray}
As those local time increase, corresponding time coordinates in an observing system (m), 
\begin{eqnarray}
\TE(s_\mathrm{e}) &\mathrm{and}& \TM(s_\mathrm{m}),
\end{eqnarray}
  also change.
On the other hand, the basic waves also advance in space coordinates in the observing system as following formulas (m), 
\begin{eqnarray}
\RE(s_\mathrm{e}) &\mathrm{and}& \RM(s_\mathrm{m}).
\end{eqnarray}
Here, we assume that the time in the observing system, $\TE$ and $\TM$, have $x$, $y$, and $z$ components as same as space, $\RE$ and $\RM$.
Moreover, we allow those coordinates to have complex values.
About definitions of operations in complex-vector space, see Appendix \ref{sec_CSdef}.
So, we assume the time and space have totally twelve dimensions (six for time and six for space).

About these assumptions, we have to explain the process how we reached at there.
In many try and errors including some references,\cite{N1,N2} real values bring us problems,
  which are divergence due to zero division, and so on.
The divergence and discontinuity in basic equations defined in following paragraphs make no solutions for us.
We believe that the nature is smooth and has no divergences and discontinuities.
So, we tried to use complex values for only the scalar fields and it seemed to go well.
However, we found that velocity vectors in space defined in following paragraphs can not avoid to have complex values in those components.
Therefore, we treat complex space, which is defined by the space including imaginary values in its coordinates.
Moreover, time also need complex value, because the basic equations have complex values.
However, complex time in one dimension having complex value easily advances along circle orbits on complex plane.
This is unphysical, because we are existing in the world with time advancing straight.
Therefore, we also treat complex time in three dimensions by the same way with complex space.

Since we expect that those basic waves interact with each other like electromagnetic waves having both electric and magnetic characters,
  we assume that those waves keep the distance from each other.
By considering those and to solve simple problems, we assume that
  the basic waves advance with the same velocities in both the space and time coordinates in observing system.
This make it possible that those local systems have the same local time at the most nearest points to each other.
Therefore, those assumptions lead to following equations,
\begin{eqnarray}
 s_\mathrm{e}&=&s_\mathrm{m}\equiv s,\\
  |\dot{\TE}|&=&|\dot{\TM}|\equiv|\dot{\VEC{t}}|,\\
  |\dot{\RE}|&=&|\dot{\RM}|\equiv|\dot{\VEC{r}}|,
\end{eqnarray}
where
\begin{eqnarray}
\dot{\TE}&\equiv& \frac{d}{ds}\TE\equiv|\dot{\TE}|\UE,\\
\dot{\TM}&\equiv& \frac{d}{ds}\TM\equiv|\dot{\TM}|\UM,
\end{eqnarray}
\begin{eqnarray}
\dot{\RE}&\equiv& \frac{d}{ds}\RE\equiv|\dot{\RE}|\EE,\\
\dot{\RM}&\equiv& \frac{d}{ds}\RM\equiv|\dot{\RM}|\EM,
\end{eqnarray}
We call $\dot{\TE}$ and $\dot{\TM}$ ($\dot{\RE}$, and $\dot{\RM}$) velocity vectors in time (space).
We also call $\UE$ and $\UM$ ($\EE$ and $\EM$) unit-velocity vectors in time (space).
Those directions are along the directions of the basic waves.
The relation among local and observing systems becomes
\begin{eqnarray}
s=
 \frac{\UE\cdot \TE}{|\dot{\VEC{t}}|}
-\frac{\EE\cdot \RE}{|\dot{\VEC{r}}|}
=
 \frac{\UM\cdot \TM}{|\dot{\VEC{t}}|}
-\frac{\EM\cdot \RM}{|\dot{\VEC{r}}|}.\label{eq_LO}
\end{eqnarray}
By substituting this relation for (\ref{eq_BW}), we find formulas of the basic waves in the observing system.
Moreover, if the local system move with the same velocity with the wave,
  then the wave does not move in the local system.
This phenomena is equivalent with that the local time in (\ref{eq_LO}) does not change.
Therefore, time advance by different speeds for different inertial systems,
  and which is described by special relativity.
We define components of local-time derivatives of coordinates in the observing system,
\begin{eqnarray}
\dot{\TE}\equiv
\left(
\begin{array}{c} \dot{\AE} \\ \dot{\BE} \\ \dot{\CE} 
\end{array}
\right)
&,&
\dot{\TM}\equiv
\left(
\begin{array}{c} \dot{\AM} \\ \dot{\BM} \\ \dot{\CM} 
\end{array}
\right)
,
\end{eqnarray}
\begin{eqnarray}
\dot{\RE}\equiv
\left(
\begin{array}{c} \dot{\XE} \\ \dot{\YE} \\ \dot{\ZE} 
\end{array}
\right)
&,&
\dot{\RM}\equiv
\left(
\begin{array}{c} \dot{\XM} \\ \dot{\YM} \\ \dot{\ZM} 
\end{array}
\right),
\end{eqnarray}
and those products with the scalar fields (V/m),
\begin{eqnarray}
\dot{\TTE}\equiv\xi\dot{\TE}\equiv
\left(
\begin{array}{c} \dot{\TAE} \\ \dot{\TBE} \\ \dot{\TCE} 
\end{array}
\right)
&,&
\dot{\TTM}\equiv\eta\dot{\TM}\equiv
\left(
\begin{array}{c} \dot{\TAM} \\ \dot{\TBM} \\ \dot{\TCM} 
\end{array}
\right)
,
\end{eqnarray}
\begin{eqnarray}
\dot{\TRE}\equiv\xi\dot{\RE}\equiv
\left(
\begin{array}{c} \dot{\TXE} \\ \dot{\TYE} \\ \dot{\TZE} 
\end{array}
\right)
&,&
\dot{\TRM}\equiv\eta\dot{\RM}\equiv
\left(
\begin{array}{c} \dot{\TXM} \\ \dot{\TYM} \\ \dot{\TZM} 
\end{array}
\right).
\end{eqnarray}

Now, we assume two energy tensors,
\begin{widetext}
\begin{eqnarray}
\ME&\equiv&
\left(
\begin{array}{c c c c c c}
   \ME^{11}                   & \ME^{12}                   & \ME^{13}                   & \ME^{14}                   & \ME^{15}                   & \ME^{16}                   \\
   \ME^{21}                   & \ME^{22}                   & \ME^{23}                   & \ME^{24}                   & \ME^{25}                   & \ME^{26}                   \\
   \ME^{31}                   & \ME^{32}                   & \ME^{33}                   & \ME^{34}                   & \ME^{35}                   & \ME^{36}                   \\
   \ME^{41}                   & \ME^{42}                   & \ME^{43}                   & \ME^{44}                   & \ME^{45}                   & \ME^{46}                   \\
   \ME^{51}                   & \ME^{52}                   & \ME^{53}                   & \ME^{54}                   & \ME^{55}                   & \ME^{56}                   \\
   \ME^{61}                   & \ME^{62}                   & \ME^{63}                   & \ME^{64}                   & \ME^{65}                   & \ME^{66}                   
\end{array}
\right)
\equiv
\left(
\begin{array}{c c c c c c}
   \dot{\TAE}^* \dot{\TAE}    & \dot{\TAE}^* \dot{\TBE}    & \dot{\TAE}^* \dot{\TCE}    & \dot{\TAE}^* \dot{\TXE}    & \dot{\TAE}^* \dot{\TYE}    & \dot{\TAE}^* \dot{\TZE}    \\
   \dot{\TBE}^* \dot{\TAE}    & \dot{\TBE}^* \dot{\TBE}    & \dot{\TBE}^* \dot{\TCE}    & \dot{\TBE}^* \dot{\TXE}    & \dot{\TBE}^* \dot{\TYE}    & \dot{\TBE}^* \dot{\TZE}    \\
   \dot{\TCE}^* \dot{\TAE}    & \dot{\TCE}^* \dot{\TBE}    & \dot{\TCE}^* \dot{\TCE}    & \dot{\TCE}^* \dot{\TXE}    & \dot{\TCE}^* \dot{\TYE}    & \dot{\TCE}^* \dot{\TZE}    \\
   \dot{\TXE}^* \dot{\TAE}    & \dot{\TXE}^* \dot{\TBE}    & \dot{\TXE}^* \dot{\TCE}    & \dot{\TXE}^* \dot{\TXE}    & \dot{\TXE}^* \dot{\TYE}    & \dot{\TXE}^* \dot{\TZE}    \\
   \dot{\TYE}^* \dot{\TAE}    & \dot{\TYE}^* \dot{\TBE}    & \dot{\TYE}^* \dot{\TCE}    & \dot{\TYE}^* \dot{\TXE}    & \dot{\TYE}^* \dot{\TYE}    & \dot{\TYE}^* \dot{\TZE}    \\
   \dot{\TZE}^* \dot{\TAE}    & \dot{\TZE}^* \dot{\TBE}    & \dot{\TZE}^* \dot{\TCE}    & \dot{\TZE}^* \dot{\TXE}    & \dot{\TZE}^* \dot{\TYE}    & \dot{\TZE}^* \dot{\TZE} 
\end{array}
\right),
\end{eqnarray}
and
\begin{eqnarray}
\MM&\equiv&
\left(
\begin{array}{c c c c c c}
   \dot{\TAM}^* \dot{\TAM}    & \dot{\TAM}^* \dot{\TBM}    & \dot{\TAM}^* \dot{\TCM}    & \dot{\TAM}^* \dot{\TXM}    & \dot{\TAM}^* \dot{\TYM}    & \dot{\TAM}^* \dot{\TZM}    \\
   \dot{\TBM}^* \dot{\TAM}    & \dot{\TBM}^* \dot{\TBM}    & \dot{\TBM}^* \dot{\TCM}    & \dot{\TBM}^* \dot{\TXM}    & \dot{\TBM}^* \dot{\TYM}    & \dot{\TBM}^* \dot{\TZM}    \\
   \dot{\TCM}^* \dot{\TAM}    & \dot{\TCM}^* \dot{\TBM}    & \dot{\TCM}^* \dot{\TCM}    & \dot{\TCM}^* \dot{\TXM}    & \dot{\TCM}^* \dot{\TYM}    & \dot{\TCM}^* \dot{\TZM}    \\
   \dot{\TXM}^* \dot{\TAM}    & \dot{\TXM}^* \dot{\TBM}    & \dot{\TXM}^* \dot{\TCM}    & \dot{\TXM}^* \dot{\TXM}    & \dot{\TXM}^* \dot{\TYM}    & \dot{\TXM}^* \dot{\TZM}    \\
   \dot{\TYM}^* \dot{\TAM}    & \dot{\TYM}^* \dot{\TBM}    & \dot{\TYM}^* \dot{\TCM}    & \dot{\TYM}^* \dot{\TXM}    & \dot{\TYM}^* \dot{\TYM}    & \dot{\TYM}^* \dot{\TZM}    \\
   \dot{\TZM}^* \dot{\TAM}    & \dot{\TZM}^* \dot{\TBM}    & \dot{\TZM}^* \dot{\TCM}    & \dot{\TZM}^* \dot{\TXM}    & \dot{\TZM}^* \dot{\TYM}    & \dot{\TZM}^* \dot{\TZM} 
\end{array}
\right),
\end{eqnarray}
and two stress tensors,
\begin{eqnarray}
\SE&\equiv&
\left(
\begin{array}{c c c c c c}
  |\dot{\TRE}|^2              & \xi^*|\dot{\RE}|\dot{\TCM} &-\xi^*|\dot{\RE}|\dot{\TBM} & \dot{\TAE} \dot{\TXE}^*    & \dot{\TAE} \dot{\TYE}^*    & \dot{\TAE} \dot{\TZE}^*    \\
  -\xi^*|\dot{\RE}|\dot{\TCM} & |\dot{\TRE}|^2             & \xi^*|\dot{\RE}|\dot{\TAM} & \dot{\TBE} \dot{\TXE}^*    & \dot{\TBE} \dot{\TYE}^*    & \dot{\TBE} \dot{\TZE}^*    \\
   \xi^*|\dot{\RE}|\dot{\TBM} &-\xi^*|\dot{\RE}|\dot{\TAM} & |\dot{\TRE}|^2             & \dot{\TCE} \dot{\TXE}^*    & \dot{\TCE} \dot{\TYE}^*    & \dot{\TCE} \dot{\TZE}^*    \\
   \dot{\TXE} \dot{\TAE}^*    & \dot{\TXE} \dot{\TBE}^*    & \dot{\TXE} \dot{\TCE}^*    & |\dot{\TTE}|^2             & \xi^*|\dot{\TE}|\dot{\TZM} &-\xi^*|\dot{\TE}|\dot{\TYM} \\
   \dot{\TYE} \dot{\TAE}^*    & \dot{\TYE} \dot{\TBE}^*    & \dot{\TYE} \dot{\TCE}^*    &-\xi^*|\dot{\TE}|\dot{\TZM} & |\dot{\TTE}|^2             & \xi^*|\dot{\TE}|\dot{\TXM} \\
   \dot{\TZE} \dot{\TAE}^*    & \dot{\TZE} \dot{\TBE}^*    & \dot{\TZE} \dot{\TCE}^*    & \xi^*|\dot{\TE}|\dot{\TYM} &-\xi^*|\dot{\TE}|\dot{\TXM} & |\dot{\TTE}|^2          
\end{array}
\right),
\end{eqnarray}
and
\begin{eqnarray}
\SM&\equiv&
\left(
\begin{array}{c c c c c c}
  |\dot{\TRM}|^2              &-\eta^*|\dot{\RM}|\dot{\TCE}& \eta^*|\dot{\RM}|\dot{\TBE}& \dot{\TAM} \dot{\TXM}^*    & \dot{\TAM} \dot{\TYM}^*    & \dot{\TAM} \dot{\TZM}^*    \\
   \eta^*|\dot{\RM}|\dot{\TCE}& |\dot{\TRM}|^2             &-\eta^*|\dot{\RM}|\dot{\TAE}& \dot{\TBM} \dot{\TXM}^*    & \dot{\TBM} \dot{\TYM}^*    & \dot{\TBM} \dot{\TZM}^*    \\
  -\eta^*|\dot{\RM}|\dot{\TBE}& \eta^*|\dot{\RM}|\dot{\TAE}& |\dot{\TRM}|^2             & \dot{\TCM} \dot{\TXM}^*    & \dot{\TCM} \dot{\TYM}^*    & \dot{\TCM} \dot{\TZM}^*    \\
   \dot{\TXM} \dot{\TAM}^*    & \dot{\TXM} \dot{\TBM}^*    & \dot{\TXM} \dot{\TCM}^*    & |\dot{\TTM}|^2             &-\eta^*|\dot{\TM}|\dot{\TZE}& \eta^*|\dot{\TM}|\dot{\TYE}\\
   \dot{\TYM} \dot{\TAM}^*    & \dot{\TYM} \dot{\TBM}^*    & \dot{\TYM} \dot{\TCM}^*    & \eta^*|\dot{\TM}|\dot{\TZE}& |\dot{\TTM}|^2             &-\eta^*|\dot{\TM}|\dot{\TXE}\\
   \dot{\TZM} \dot{\TAM}^*    & \dot{\TZM} \dot{\TBM}^*    & \dot{\TZM} \dot{\TCM}^*    &-\eta^*|\dot{\TM}|\dot{\TYE}& \eta^*|\dot{\TM}|\dot{\TXE}& |\dot{\TTM}|^2          
\end{array}
\right).
\end{eqnarray}

For also these assumptions, we have to explain the process how we got them.
In many try and errors including some references,\cite{N1,N2} we easily lost solutions in the basic equations.
That is mainly because of cancellations between left and right hand sides of the basic equations.
We believe that the nature does not need what it does not need.
Therefore, we put complex terms into above matrices not to occur any cancellations.
Moreover, we also put terms to make matrices have high symmetry between space and time,
  because we believe that the nature prefers high symmetry.
Furthermore, we also put terms to make the basic equations satisfy Maxwell equations in some conditions 
  as explained in following sections.
We note that there are still ambiguities in determining terms in these matrices.
For example, changing all rotation parts as shown in following paragraphs has a possible to be assumed,
  but which does not make any differences in qualitative discussions.

Then, we construct the basic equations as
\begin{eqnarray}
\partial_\nu \ME^{\mu\nu}&=& \partial_\nu \SE^{\mu\nu},\label{eq_BE1a}\\
\partial_\nu \MM^{\mu\nu}&=& \partial_\nu \SM^{\mu\nu},\label{eq_BE1b}
\end{eqnarray}
where
\begin{eqnarray}
\partial_\nu \ME^{1\sim3,\nu}
&=&
 \left(\dot{\TTE}^*\cdot\nabla_t\right)\dot{\TTE}^* + \dot{\TTE}^*\left(\nabla_t \cdot \dot{\TTE} \right)
+\left(\dot{\TRE}^*\cdot\nabla  \right)\dot{\TTE}^* + \dot{\TTE}^*\left(\nabla   \cdot \dot{\TRE} \right)
,\label{eq_BE3a}\\
\partial_\nu \ME^{4\sim6,\nu}
&=&
 \left(\dot{\TTE}^*\cdot\nabla_t\right)\dot{\TRE}^* + \dot{\TRE}^*\left(\nabla_t \cdot \dot{\TTE} \right)
+\left(\dot{\TRE}^*\cdot\nabla  \right)\dot{\TRE}^* + \dot{\TRE}^*\left(\nabla   \cdot \dot{\TRE} \right)
,\label{eq_BE3b}\\
\partial_\nu \SE^{1\sim3,\nu}
&=&
 \nabla_t |\dot{\TRE}|^2 + \nabla_t\times\left(\xi|\dot{\RE}|\dot{\TTM}^*\right)
+\left(\dot{\TRE}\cdot\nabla  \right)\dot{\TTE} + \dot{\TTE}\left(\nabla   \cdot \dot{\TRE}^* \right)
,\label{eq_BE3c}\\
\partial_\nu \SE^{4\sim6,\nu}
&=&
 \left(\dot{\TTE}\cdot\nabla_t\right)\dot{\TRE} + \dot{\TRE}\left(\nabla_t \cdot \dot{\TTE}^* \right)
+\nabla   |\dot{\TTE}|^2 + \nabla\times\left(\xi|\dot{\TE}|\dot{\TRM}^*\right)
,\label{eq_BE3d}
\end{eqnarray}
and $\nabla_t$ means a gradation operator for time coordinate in the observing system.
For $\partial_\nu \MM^{\mu\nu}$ and $\partial_\nu \SM^{\mu\nu}$, 
  we only have to exchange ($\xi$, ${}_\mathrm{e}$) for ($\eta$, ${}_\mathrm{m}$) 
  and change signs of rotation parts.
To find solutions of the basic equations easily, we consider steady states.
Therefore, we assume
\begin{eqnarray}
\frac{d}{ds} |\dot{\VEC{r}}|=
\frac{d}{ds} |\dot{\VEC{t}}|=0,
\end{eqnarray}
  which means that the basic waves have constant speeds in both local and observing systems.
By using this relation, the basic equations become simple as following equations,
\begin{eqnarray}
& &\dot{\TUE}^*\left(\TUE^*\cdot\UE-\TEE^*\cdot\EE\right)
   +\TUE^*\left(\UE\cdot\dot{\TUE}-\EE\cdot\dot{\TEE}\right) \nonumber\\
&=&
+\beta^2\UE\frac{d}{ds}|\xi|^2
+\beta\left( \UE\dot{\xi}^*\times\TUM^*+\xi^*\UM\times\dot{\TUM}^*\right)
-\dot{\TUE}\left(\TEE\cdot\EE\right)-\TUE\left(\EE\cdot\dot{\TEE}^*\right),\label{eq_BE5a}\\
& &\dot{\TEE}^*\left(\TUE^*\cdot\UE-\TEE^*\cdot\EE\right)
   +\TEE^*\left(\UE\cdot\dot{\TUE}-\EE\cdot\dot{\TEE}\right) \nonumber\\
&=&
-\frac{1}{\beta^2}\EE\frac{d}{ds}|\xi|^2
-\frac{1}{\beta}\left( \EE\dot{\xi}^*\times\TEM^*+\xi^*\EM\times\dot{\TEM}^*\right)
+\dot{\TEE}\left(\TUE\cdot\UE\right)+\TEE\left(\UE\cdot\dot{\TUE}^*\right)
,\label{eq_BE5b}
\end{eqnarray}
\end{widetext}
where
\begin{eqnarray}
\TUE\equiv\xi\UE&,&\TUM\equiv\xi\UM,\\
\TEE\equiv\xi\EE&,&\TEM\equiv\xi\EM,
\end{eqnarray}
\begin{eqnarray}
\dot{\TUE}\equiv\frac{d}{ds}\TUE&,&\dot{\TUM}\equiv\frac{d}{ds}\TUM,\\
\dot{\TEE}\equiv\frac{d}{ds}\TEE&,&\dot{\TEM}\equiv\frac{d}{ds}\TEM,
\end{eqnarray}
\begin{eqnarray}
\dot{\UE}\equiv\frac{d}{ds}\UE&,&\dot{\UM}\equiv\frac{d}{ds}\UM,\\
\dot{\EE}\equiv\frac{d}{ds}\EE&,&\dot{\EM}\equiv\frac{d}{ds}\EM,
\end{eqnarray}
\begin{eqnarray}
\ddot{\TTE} \equiv  \frac{d}{ds} \dot{\TTE} = |\dot{\VEC{t}}| \dot{\TUE}
  &,&
\ddot{\TTM} \equiv  \frac{d}{ds} \dot{\TTM} = |\dot{\VEC{t}}| \dot{\TUM},\\
\ddot{\TRE} \equiv  \frac{d}{ds} \dot{\TRE} = |\dot{\VEC{r}}| \dot{\TEE}
  &,&
\ddot{\TRM} \equiv  \frac{d}{ds} \dot{\TRM} = |\dot{\VEC{r}}| \dot{\TEM},
\end{eqnarray}
\begin{eqnarray}
\ddot{\TE} \equiv  \frac{d}{ds} \dot{\TE} = |\dot{\VEC{t}}| \dot{\UE}
  &,&
\ddot{\TM} \equiv  \frac{d}{ds} \dot{\TM} = |\dot{\VEC{t}}| \dot{\UM},\\
\ddot{\RE} \equiv  \frac{d}{ds} \dot{\RE} = |\dot{\VEC{r}}| \dot{\EE}
  &,&
\ddot{\RM} \equiv  \frac{d}{ds} \dot{\RM} = |\dot{\VEC{r}}| \dot{\EM},
\end{eqnarray}
\begin{eqnarray}
\dot{\xi}\equiv\frac{d}{ds}\xi&,&\dot{\eta}\equiv\frac{d}{ds}\eta,\\
\beta&\equiv&\frac{|\dot{\VEC{r}}|}{|\dot{\VEC{t}}|},
\end{eqnarray}
  and, about detail operations in complex-vector space, see Appendix \ref{sec_CSapp}.
About corresponding equations for (\ref{eq_BE1b}), we only have to exchange ($\xi$, ${}_\mathrm{e}$)
  for ($\eta$, ${}_\mathrm{m}$) and change signs of rotation parts.
We call $\ddot{\TE}$ and $\ddot{\TM}$ ($\ddot{\RE}$ and $\ddot{\RM}$) acceleration vectors in time (space).
We also call $\dot{\UE}$ and $\dot{\UM}$ ($\dot{\EE}$ and $\dot{\EM}$) acceleration-unit vectors in time (space).
We note that the acceleration-unit vector does not promise unit length,
  although the unit-velocity vector has always unit length.
We also note that time and space are equivalent with each other in these basic equations,
  because exchanging ($\dot{\TE}$, $\dot{\TM}$) for ($\dot{\RE}$, $\dot{\RM}$) derives the same equations.
Since time and space are equivalent in this theory,
  there is a possibility of existing another world,
  where our time (space) works as their space (time).

\subsection{Lorentz force and Maxwell equations}
On the two basic waves, the most nearest points to each other are expected to exist in the same time in observing system,
  because those need to interact with each other.
Moreover, the waves are expected to advance with an constant speed in time coordinates like 
  many waves in steady states in our world.
Therefore, we assume that the two local-time derivatives of time in observing system 
  are equal to each other and those local-time derivatives are zero
  as following relations,
\begin{eqnarray}
        \UE=\UM      &\equiv&\VEC{u},\label{eq_UCONST1}\\
  \dot{\UE}=\dot{\UM}&  =   &0.      \label{eq_UCONST2}
\end{eqnarray}
Then, we define charges (V/m$^2$), currents (V/m$^2$), and fields (V/m) as following formulas,
\begin{eqnarray}
  \rho_\mathrm{e}\equiv     \dot{\xi}                                         &,&\rho_\mathrm{m}\equiv -   \dot{\eta},\\
     j_\mathrm{e}\equiv -   \nabla'\xi=\frac{1}{\beta}\EE\dot{\xi}            &,&   j_\mathrm{m}\equiv +   \nabla'\eta=-\frac{1}{\beta}\EM\dot{\eta},\\
  E\equiv -   \xi\frac{j_\mathrm{e}}{\rho_\mathrm{e}}=-\frac{1}{\beta}\TEE    &,&B\equiv +   \eta\frac{j_\mathrm{m}}{\rho_\mathrm{m}}=+\frac{1}{\beta}\TEM,
\end{eqnarray}
where $\nabla'$ is a gradation operator in the local systems,
\begin{eqnarray}
\nabla'\equiv |\dot{\VEC{t}}|\nabla.
\end{eqnarray}
This is because that space coordinates in the local system, $\RE'$ and $\RM'$, are scaled by the factor $|\dot{\VEC{t}}|$,
  and which explain that the wave velocity, $\beta$, is the same between the local and observing systems,
\begin{eqnarray}
\beta=\frac{|\dot{\VEC{r}}|}{|\dot{\VEC{t}}|}
     =\left|\frac{d\RE'}{ds}\right|
     =\left|\frac{d\RM'}{ds}\right|
     =\frac{d}{d|\VEC{t}|}|\RE|
     =\frac{d}{d|\VEC{t}|}|\RM|.
\end{eqnarray}
This phenomena is also shown in special relativity.
To reduce difficulties of finding solutions, as the basic waves, we assume
\begin{eqnarray}
  \xi&=& \xi_0 \mathrm{e}^{i\omega s},       \label{eq_SCONST1}\\
 \eta&=& \eta_0 \mathrm{e}^{i\omega s}.      \label{eq_SCONST2}
\end{eqnarray}
Since electromagnetic waves advance straight in real space,
  which is defined by the space having non-imaginary values in its coordinates,
  we consider that the basic waves advance in real space.
So, both $\EE$ and $\EM$ have real values in those components as explained in the following formula,
\begin{eqnarray}
  \EE\cdot\EE^*&=&\EM\cdot\EM^*=+1.\label{eq_CEEP}
\end{eqnarray}
We also choose zero or $\pi$ for the phase difference between the two basic waves,
  and which is shown as the following relation,
\begin{eqnarray}
\alpha^*&=&\alpha,\label{eq_PHASE}
\end{eqnarray}
where
\begin{eqnarray}
\alpha&\equiv&\frac{\xi_0}{\eta_0}.
\end{eqnarray}
Therefore, in this case, we find following relations,
\begin{eqnarray}
  \frac{d}{ds}|\xi|^2 &=& \dot{\xi }^*\xi +\xi ^*\dot{\xi }=0,\\
  \frac{d}{ds}|\eta|^2&=& \dot{\eta}^*\eta+\eta^*\dot{\eta}=0,\\
                      & & \dot{\xi }^*\eta+\eta^*\dot{\xi }=0,\\
                      & & \dot{\eta}^*\xi +\xi ^*\dot{\eta}=0,
\end{eqnarray}
which also lead to following relations,
\begin{eqnarray}
    \xi\rho_\mathrm{e}^*=-\xi^*\rho_\mathrm{e} &,&   \eta\rho_\mathrm{m}^*=-\eta^*\rho_\mathrm{m},\label{eq_Charge}\\
      \xi j_\mathrm{e}^*=-\xi^*j_\mathrm{e}    &,&     \eta j_\mathrm{m}^*=-\eta^*j_\mathrm{m},\\
j_\mathrm{e}^*\times B^*=-j_\mathrm{e}\times B &,&j_\mathrm{m}^*\times E^*=-j_\mathrm{m}\times E.
\end{eqnarray}
By using above definitions and relations, from the right hand sides of (\ref{eq_BE5a}), (\ref{eq_BE5b}), 
  and the corresponding equations for (\ref{eq_BE1b}), we get following formulas,
\begin{widetext}
\begin{eqnarray}
  \partial_\nu \SE^{1\sim3,\nu}
    &=&+\xi\VEC{u}\left(\rho_\mathrm{e}^*-\beta^2 \nabla'\cdot E^*\right),\\
  \partial_\nu \SE^{4\sim6,\nu}
    &=& \beta\left[ E\rho_\mathrm{e}^*\left\{ \frac{1}{\beta^2}-\left(\VEC{u}\cdot\VEC{u}^*\right)\right\} + j_\mathrm{e}\times B \right]
       -\beta  \xi^*\left[ \dot{E}  -\nabla'\times B^*+\frac{1}{\beta^2}j_\mathrm{e}  \right],\label{eq_force1}\\
  \partial_\nu \SM^{1\sim3,\nu}
    &=&-\eta\VEC{u}\left(\rho_\mathrm{m}^*-\beta^2 \nabla'\cdot B^*\right),\\
  \partial_\nu \SM^{4\sim6,\nu}
    &=& \beta\left[ B\rho_\mathrm{m}^*\left\{ \frac{1}{\beta^2}-\left(\VEC{u}\cdot\VEC{u}^*\right)\right\} - j_\mathrm{m}\times E \right]
       +\beta \eta^*\left[ \dot{B}  +\nabla'\times E^*+\frac{1}{\beta^2}j_\mathrm{m}  \right],\label{eq_force2}
\end{eqnarray}
\end{widetext}
where
\begin{eqnarray}
  \dot{E}&\equiv&\frac{d}{ds}E,\\
  \dot{B}&\equiv&\frac{d}{ds}B.
\end{eqnarray}
If we assume
\begin{eqnarray}
\beta&=&1,\label{eq_BETA1}\\
\VEC{u}\cdot\VEC{u}^*&=&+1,\label{eq_UUREAL}\\
\end{eqnarray}
  then we find well known terms in Maxwell equations and Lorentz force.
Here, the relation of (\ref{eq_BETA1}) means that the wave velocity is light speed.
From the coefficients of first force terms in (\ref{eq_force1}) and (\ref{eq_force2}),
\begin{eqnarray}
     \frac{1}{\beta^2}-\left(\VEC{u}\cdot\VEC{u}^*\right),
\end{eqnarray}
  the wave velocities seem to be controlled to be light speed in the condition of (\ref{eq_UUREAL}).
The relation, (\ref{eq_UUREAL}), means that the wave advance in real time,
  which is defined by the time having non-imaginary values in its coordinates.
Those relations leading to the terms in Lorentz force and Maxwell equations,
  (\ref{eq_CEEP}), (\ref{eq_PHASE}), (\ref{eq_BETA1}), and (\ref{eq_UUREAL}), do not lose solutions,
  and which are expected to have electromagnetic waves near themselves as explained in following sections.
Moreover, the equations
\begin{eqnarray}
  \partial_\nu \SE^{\mu\nu}&=&0,\\
  \partial_\nu \SM^{\mu\nu}&=&0,
\end{eqnarray}
mean the conditions giving solutions without forces in real space.
On the other hand, in complex space,
  the waves are affected by other forces and can not advance straight as explained in following sections.

\section{ANALYSIS}
\subsection{Waves moving in real space}
Let us go back (\ref{eq_BE5a}) and (\ref{eq_BE5b}),
  and we consider that the basic waves advance in both real time and real space,
  which leads to the following relation,
\begin{eqnarray}
  \UE\cdot\UE^*&=&\UM\cdot\UM^*=+1,\label{eq_CUU}
\end{eqnarray}
  and (\ref{eq_CEEP}).
Those relations mean that the unit-velocity vectors in both time and space, $\UE$, $\UM$, $\EE$, and $\EM$, give real values in those components.
We assume those conditions to find an solution of electromagnetic waves,
  because the electromagnetic waves advance straight in real space.
Moreover, we assume (\ref{eq_SCONST1}) and (\ref{eq_SCONST2}) again to reduce the difficulties of finding solutions.
Then, (\ref{eq_BE5a}) and (\ref{eq_BE5b}) become
\begin{widetext}
\begin{eqnarray}
-2\UE\delta^-_\mathrm{e}\omega
  +\dot{\UE}+\UE\left(\UE\cdot\dot{\UE}\right)
&=&+\frac{\beta}{\alpha}\left(-i\omega \UE\times\UM^* + \UM\times\dot{\UM}^* \right),\label{eq_BEaa}\\
+2\EE\gamma^-_\mathrm{e}  \omega
  -\dot{\EE}-\EE\left(\EE\cdot\dot{\EE}\right)
&=&-\frac{1}{\alpha\beta}\left(-i\omega \EE\times\EM^* + \EM \times \dot{\EM}^*  \right),\label{eq_BEab}
\end{eqnarray}
\end{widetext}
where
\begin{eqnarray}
  \delta^\pm_\mathrm{e}\equiv\frac{\EE\cdot\left(\dot{\EE}\pm\dot{\EE}^*\right)}{2\omega}&,&
  \delta^\pm_\mathrm{m}\equiv\frac{\EM\cdot\left(\dot{\EM}\pm\dot{\EM}^*\right)}{2\omega},\\
  \gamma^\pm_\mathrm{e}\equiv\frac{\UE\cdot\left(\dot{\UE}\pm\dot{\UE}^*\right)}{2\omega}&,&
  \gamma^\pm_\mathrm{m}\equiv\frac{\UM\cdot\left(\dot{\UM}\pm\dot{\UM}^*\right)}{2\omega},
\end{eqnarray}
About corresponding equations for (\ref{eq_BE1b}), we only have to exchange ($\xi$, ${}_\mathrm{e}$)
  for ($\eta$, ${}_\mathrm{m}$) and change signs of rotation parts.
Furthermore, we realistically assume again that the waves advance in real time straight with a constant speed as explained in (\ref{eq_UCONST1}) and (\ref{eq_UCONST2}).
To satisfy (\ref{eq_BEaa}) and the corresponding equation for (\ref{eq_BE1b}), those assumptions lead to
\begin{eqnarray}
  \delta^-_\mathrm{e}=\delta^-_\mathrm{m}=\gamma^\pm_\mathrm{e}=\gamma^\pm_\mathrm{m}=0.\label{eq_DG1}
\end{eqnarray}
Therefore, $\dot{\EE}$ and $\dot{\EM}$ ($\dot{\UE}$ and $\dot{\UM}$) give real (zero) values on $\EE$ and $\EM$ ($\UE$ and $\UM$), respectively.

\begin{figure}[htbp]
  \vspace{0cm}
  \begin{center}
  \includegraphics[width=0.25\textwidth]{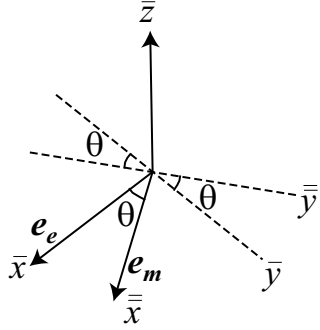}
  \caption{
    Unit-velocity vectors in space, $\EE$ and $\EM$, having an angle, $\theta$, 
      between themselves.
    Those unit-velocity vectors, $\EE$ and $\EM$, are related to electric and magnetic fields, respectively.
    Coordinates, $\bar{x}$, $\dbar{x}$, $\bar{y}$, $\dbar{y}$, and $\bar{z}$ are local coordinates.
  }
  \label{fig_10}
  \end{center}
\end{figure}

Now, we consider that there are an angle of $\theta$ between the unit-velocity vectors in space, $\EE$ and $\EM$, as shown in Fig. \ref{fig_10}.
Those unit-velocity vectors, $\EE$ and $\EM$, are along the $\bar{x}$ and $\dbar{x}$ axis, respectively.
Then, we write down (\ref{eq_BEab}) and the corresponding equation for (\ref{eq_BE1b}) in terms of each component in space.
Therefore, those equations become
\begin{eqnarray}
  -2\DEEX&=&-\frac{1}{\alpha\beta}\left[+\EMX\DEMZ\sin\theta \right],\label{eq_BEba}\\
   -\DEEY&=&-\frac{1}{\alpha\beta}\left[-\EMX\DEMZ\cos\theta \right],\label{eq_BEbb}\\
   -\DEEZ&=&-\frac{1}{\alpha\beta}\left[-i\omega\EEX\EMX\sin\theta+\EMX\DEMY \right],\label{eq_BEbc}
\end{eqnarray}
\begin{eqnarray}
  -2\DEMX&=&-\frac{\alpha}{\beta} \left[+\EEX\DEEZ\sin\theta \right],\label{eq_BEbd}\\
   -\DEMY&=&-\frac{\alpha}{\beta} \left[+\EEX\DEEZ\cos\theta \right],\label{eq_BEbe}\\
   -\DEMZ&=&-\frac{\alpha}{\beta} \left[-i\omega\EMX\EEX\sin\theta-\EEX\DEEY \right].\label{eq_BEbf}
\end{eqnarray}
Here, we choose $\EEX$ and $\EMX$ as following formulas,
\begin{eqnarray}
  \EEX=\EMX&=& +1, \label{eq_CEE}
\end{eqnarray}
  because the unit-velocity vectors in space give real values in those components as shown in (\ref{eq_CEEP}).
We also choose $\alpha$ as following formulas,
\begin{eqnarray}
     \alpha&=&\pm 1,\ \mathrm{or}\ \pm i,\label{eq_CALP}
\end{eqnarray}
  which means that we search solutions in the conditions of the two basic waves' phase differences of zero, $\pm\frac{\pi}{2}$, and $\pi$.
Then, without considering (\ref{eq_DG1}), we get following solutions,
\begin{eqnarray}
  \DEEZ=&-\frac{i}{\alpha}\omega 
    \left[ \frac{\frac{1}{\beta}\sin\theta}{1-\frac{1}{\beta^2}\cos\theta} \right]&=+\frac{ \DEMZ }{\alpha^2},\\
  \DEEY=& i \omega  \frac{\cos\theta}{\beta}
    \left[ \frac{\frac{1}{\beta}\sin\theta}{1-\frac{1}{\beta^2}\cos\theta} \right]&=-\DEMY,\\
  \DEEX=&-i \omega
    \frac{\sin\theta}{2\beta} \left[\frac{\frac{1}{\beta}\sin\theta}{1-\frac{1}{\beta^2}\cos\theta} \right]&=+\DEMX.
\end{eqnarray}
From those solutions, we find an solution satisfying (\ref{eq_DG1}) from the case of
\begin{eqnarray}
  \theta=\pi,\label{eq_THETA}
\end{eqnarray}
  which leads to
\begin{eqnarray}
  \DEE=\DEM=0.
\end{eqnarray}
We notice that, from (\ref{eq_CEE}) and (\ref{eq_THETA}), the directions of the two waves are opposite to each other.
About solutions describing electromagnetic waves, we hope that those are existing in other range of $\theta$,
\begin{eqnarray}
0 < \theta < \pi,
\end{eqnarray}
However, in this range, the conditions of (\ref{eq_DG1}) are broken 
  due to imaginary values of $\DEEX$ and $\DEMX$.
Therefore, the electromagnetic waves are expected to break the realistic assumptions, (\ref{eq_UCONST1}) and (\ref{eq_UCONST2}).

\subsection{Waves moving in complex space}
Let us go back (\ref{eq_BE5a}) and (\ref{eq_BE5b}) again.
On the contrary to the previous section,
  we consider that the basic waves advance in both real time and complex space,
  so that we assume (\ref{eq_CUU}) and 
\begin{eqnarray}
  \EE\cdot\EE^*&=&\EM\cdot\EM^*=-1.\label{eq_CEEM}
\end{eqnarray}
Those relations mean that the unit-velocity vectors in time (space), $\UE$ and $\UM$ ($\EE$ and $\EM$), give real (imaginary) values in those components.
Moreover, we assume (\ref{eq_SCONST1}) and (\ref{eq_SCONST2}) again to simplify problems.
Then, (\ref{eq_BE5a}) and (\ref{eq_BE5b}) become
\begin{widetext}
\begin{eqnarray}
  -2\UE\delta^-_\mathrm{e}\omega
    +2\dot{\UE}^*+\dot{\UE}+\UE\left(\UE\cdot\dot{\UE}\right)
    &=&
    +\frac{\beta}{\alpha}\left(-i\omega \UE\times\UM^* + \UM\times\dot{\UM}^*\right),   \label{eq_BE7a}\\
  -2\EE^*\left(i-\gamma^+_\mathrm{e}\right)\omega
    +2\dot{\EE}^*-\dot{\EE}-\EE^*\left(\EE\cdot\dot{\EE}\right)
    &=&
    -\frac{1}{\alpha\beta}\left(-i\omega \EE\times\EM^* + \EM \times \dot{\EM}^*\right).\label{eq_BE7b}
\end{eqnarray}
\end{widetext}
About corresponding equations for (\ref{eq_BE1b}), we only have to exchange ($\xi$, ${}_\mathrm{e}$)
  for ($\eta$, ${}_\mathrm{m}$) and change signs of rotation parts.
Furthermore, we realistically assume that the waves advance in real time straight with a constant speed, (\ref{eq_UCONST1}) and (\ref{eq_UCONST2}), again.
To satisfy (\ref{eq_BE7a}) and the corresponding equation for (\ref{eq_BE1b}), those assumptions lead to (\ref{eq_DG1}) again.
As same as the previous section, from (\ref{eq_BE7b}) and the corresponding equation for (\ref{eq_BE1b}), we finally get following equations,
\begin{eqnarray}
  -2i\omega \EEX^* +2\DEEX^*&=&-\frac{1}{\alpha\beta}\left[+\EMX\DEMZ\sin\theta \right],\label{eq_BE9a}\\
             -\DEEY+2\DEEY^*&=&-\frac{1}{\alpha\beta}\left[-\EMX\DEMZ\cos\theta \right],\label{eq_BE9b}\\
             -\DEEZ+2\DEEZ^*&=&-\frac{1}{\alpha\beta}\left[-i\omega\EEX\EMX\sin\theta+\EMX\DEMY \right],\nonumber\\
  & &\label{eq_BE9c}
\end{eqnarray}
\begin{eqnarray}
  -2i\omega \EMX^* +2\DEMX^*&=&-\frac{\alpha}{\beta} \left[+\EEX\DEEZ\sin\theta \right],\label{eq_BE9d}\\
             -\DEMY+2\DEMY^*&=&-\frac{\alpha}{\beta} \left[+\EEX\DEEZ\cos\theta \right],\label{eq_BE9e}\\
             -\DEMZ+2\DEMZ^*&=&-\frac{\alpha}{\beta} \left[-i\omega\EMX\EEX\sin\theta-\EEX\DEEY \right].\nonumber\\
  & &\label{eq_BE9f}
\end{eqnarray}
Here, we choose $\EEX$ and $\EMX$ as following formulas,
\begin{eqnarray}
  \EEX=\EMX&=& +i,
\end{eqnarray}
  because the unit-velocity vectors in space give imaginary values in those components as shown in (\ref{eq_CEEM}).
We choose $\alpha$ as (\ref{eq_CALP}) again and search solutions.
Since the conditions, (\ref{eq_DG1}), show that the components along the velocity vectors for acceleration-unit vectors in space, $\DEEX$ and $\DEMX$, are real values, 
  the acceleration-unit vectors, $\DEE$ and $\DEM$, give  real values in all components for the case of $\alpha=\pm i$.
On the other hand, the case of $\alpha=\pm 1$ leads that the components, $\DEEZ$ and $\DEMZ$, are imaginary values
  and the other components are real values.
By considering those, we solve the equations, (\ref{eq_BE9a})$\sim$(\ref{eq_BE9f}),  and get solutions as following formulas,
\begin{eqnarray}
\DEEZ=&-\frac{i}{\alpha}\omega 
 \left[ \frac{\frac{1}{\beta}\sin\theta}{-1-2\alpha^2+\frac{1}{\beta^2}\cos\theta} \right]&=+\frac{\DEMZ}{\alpha^2},\\
\DEEY=& \omega  \frac{\cos\theta}{\beta}
 \left[ \frac{\frac{1}{\beta}\sin\theta}{-1-2\alpha^2+\frac{1}{\beta^2}\cos\theta} \right]&=-\DEMY,\\
\DEEX=& \omega
 \left[ 1- \frac{\sin\theta}{2\beta} \frac{\frac{1}{\beta}\sin\theta}{-1-2\alpha^2+\frac{1}{\beta^2}\cos\theta} \right]&=+\DEMX.
\end{eqnarray}
From the signs of the components, $\DEEY$ and $\DEMY$, 
  we understand that the angle $\theta$ is kept at zero ($\frac{\pi}{2}$) for $\alpha=\pm i$ ($\pm 1$),
  if the angle gives an steady state.
Then, we find an steady-state solution from
\begin{eqnarray}
\theta=0&,&\alpha=\pm i,
\end{eqnarray}
which leads to
\begin{eqnarray}
  \DEEZ=\DEMZ=\DEEY=\DEMY&=&0,\\
              \DEEX=\DEMX&=&\omega.
\end{eqnarray}
In this solution, the two basic waves advance the same orbit, which is an circle with an radius, $\frac{\beta}{\omega}$, in the complex space like an closed string.
Since the angular velocity in space, $\omega$, is equal to the frequency in time, the closed string seems an steady state.
Moreover, the steady states have phase differences of $\pm \frac{\pi}{2}$ between the two basic waves.
We believe that more detailed analysis clear physical meanings of this solution.

\section{SUMMARY}
We proposed the theory in electromagnetic dynamics, 
  in which time and space are equivalent with each other and have totally twelve dimensions.
Then, we solved that with realistic assumptions and found a steady states as a solution.
The solution draws a circle in complex space as a closed string,
  which is an orbit passed by the waves of the electroscalar and magnetoscalar fields.
We also discussed the possibility of existing solutions of electromagnetic waves.

\appendix
\section{Complex vector space}\label{sec_CS}
\subsection{Definitions}\label{sec_CSdef}
In complex vector space, we define innerproduct,
\begin{eqnarray}
\VEC{A}\cdot\VEC{B}
&\equiv&
\sum_{i} \VEC{A}^*_i\VEC{B}_i,\\
\VEC{A}\cdot\VEC{B}&=&(\VEC{B}\cdot\VEC{A})^*,
\end{eqnarray}
 outerproduct,
\begin{eqnarray}
[\VEC{A}\times\VEC{B}]_i
&\equiv&
\VEC{A}_j\VEC{B}^*_k - \VEC{A}_k\VEC{B}^*_j,\\
\VEC{A}\times\VEC{B}&=&-(\VEC{B}\times\VEC{A})^*,
\end{eqnarray}
and norm,
\begin{eqnarray}
|\VEC{A}| &\equiv& \sqrt{\VEC{A}\cdot\VEC{A}},
\end{eqnarray}
where
\begin{eqnarray}
(ijk)=(xyz),\ (yzx),\ \mathrm{or}\ (zxy).
\end{eqnarray}

\subsection{Applications}\label{sec_CSapp}
To derive (\ref{eq_BE5a}) and (\ref{eq_BE5b}) from (\ref{eq_BE1a}),
  we need to arrange formulas appeared in (\ref{eq_BE3a})$\sim$(\ref{eq_BE3d}).
The details are following formulas,
\begin{eqnarray}
 \left(\dot{\TTE}^*\cdot\nabla_t \right) \dot{\TTE}^*
  &=& 
  \left(\dot{\TTE}^*\cdot\UE\right) \frac{\ddot{\TTE}^*}{|\dot{\VEC{t}}|}
   =
  \left(\TUE^*\cdot\UE\right) \ddot{\TTE}^*,\\
 \dot{\TTE}^* \left(\nabla_t \cdot \dot{\TTE}  \right)
  &=& 
  \dot{\TTE}^*\left(\UE\cdot\frac{\ddot{\TTE}}{|\dot{\VEC{t}}|} \right)
   =
  \dot{\TTE}^*\left(\UE\cdot\dot{\TUE} \right),
\end{eqnarray}

\begin{eqnarray}
-\left(\dot{\TRE}^*\cdot\nabla   \right) \dot{\TTE}^*
  &=& 
  \left(\dot{\TRE}^*\cdot\EE\right) \frac{\ddot{\TTE}^*}{|\dot{\VEC{r}}|}
   =
  \left(\TEE^*\cdot\EE\right) \ddot{\TTE}^*,\\
-\dot{\TTE}^* \left(\nabla   \cdot \dot{\TRE}  \right)
  &=& 
  \dot{\TTE}^*\left(\EE\cdot\frac{\ddot{\TRE}}{|\dot{\VEC{r}}|} \right)
   =
  \dot{\TTE}^*\left(\EE\cdot\dot{\TEE} \right),
\end{eqnarray}

\begin{eqnarray}
 \left(\dot{\TTE}^*\cdot\nabla_t \right) \dot{\TRE}^*
  &=& 
  \left(\dot{\TTE}^*\cdot\UE\right) \frac{\ddot{\TRE}^*}{|\dot{\VEC{t}}|}
   =
  \left(\TUE^*\cdot\UE\right) \ddot{\TRE}^*,\\
 \dot{\TRE}^* \left(\nabla_t \cdot \dot{\TTE}  \right)
  &=& 
  \dot{\TRE}^*\left(\UE\cdot\frac{\ddot{\TTE}}{|\dot{\VEC{t}}|} \right)
   =
  \dot{\TRE}^*\left(\UE\cdot\dot{\TUE} \right),
\end{eqnarray}

\begin{eqnarray}
-\left(\dot{\TRE}^*\cdot\nabla   \right) \dot{\TRE}^*
  &=& 
  \left(\dot{\TRE}^*\cdot\EE\right) \frac{\ddot{\TRE}^*}{|\dot{\VEC{r}}|}
   =
  \left(\TEE^*\cdot\EE\right) \ddot{\TRE}^*,\\
-\dot{\TRE}^* \left(\nabla   \cdot \dot{\TRE}  \right)
  &=& 
  \dot{\TRE}^*\left(\EE\cdot\frac{\ddot{\TRE}}{|\dot{\VEC{r}}|} \right)
   =
  \dot{\TRE}^*\left(\EE\cdot\dot{\TEE} \right),
\end{eqnarray}

\begin{eqnarray}
-\left(\dot{\TRE}  \cdot\nabla   \right) \dot{\TTE}  
  &=& 
  \left(\dot{\TRE}  \cdot\EE\right) \frac{\ddot{\TTE}  }{|\dot{\VEC{r}}|}
   =
  \left(\TEE  \cdot\EE\right) \ddot{\TTE}  ,\\
-\dot{\TTE}   \left(\nabla   \cdot \dot{\TRE}^*\right)
  &=& 
  \dot{\TTE}  \left(\EE\cdot\frac{\ddot{\TRE}^*}{|\dot{\VEC{r}}|} \right)
   =
  \dot{\TTE}  \left(\EE\cdot\dot{\TEE}^* \right),
\end{eqnarray}

\begin{eqnarray}
 \left(\dot{\TTE}  \cdot\nabla_t \right) \dot{\TRE}  
  &=& 
  \left(\dot{\TTE}  \cdot\UE\right) \frac{\ddot{\TRE}  }{|\dot{\VEC{t}}|}
   =
  \left(\TUE  \cdot\UE\right) \ddot{\TRE}  ,\\
 \dot{\TRE}   \left(\nabla_t \cdot \dot{\TTE}^*\right)
  &=& 
  \dot{\TRE}  \left(\UE\cdot\frac{\ddot{\TTE}^*}{|\dot{\VEC{t}}|} \right)
   =
  \dot{\TRE}  \left(\UE\cdot\dot{\TUE}^* \right),
\end{eqnarray}

\begin{eqnarray}
  \nabla_t |\dot{\TRE}|^2 
  &=&
  |\dot{\VEC{r}}|^2\nabla_t |\xi|^2 
   =
 |\dot{\VEC{r}}|^2 \UE \frac{ \frac{d}{ds} |\xi|^2 } {|\dot{t}|},
\end{eqnarray}

\begin{eqnarray}
  \nabla_t \times\left( \xi|\dot{\VEC{r}}|\dot{\TTM}^* \right)
  &=&
 |\dot{\VEC{r}}| \left[ \nabla_t \xi^* \times \dot{\TTM}^* + \xi^* \nabla_t \times \dot{\TTM}^* \right]\nonumber\\
 &=&
 |\dot{\VEC{r}}| \left[ \UE \frac{\dot{\xi}^*}{|\dot{\VEC{t}}|} \times \dot{\TTM}^* + \xi^* \UM \times \frac{\ddot{\TTM}^*}{|\dot{\VEC{t}}|} \right]\nonumber\\
 &=&
 |\dot{\VEC{r}}| \left[ \UE \dot{\xi}^* \times \TUM^* + \xi^* \UM \times \dot{\TUM}^* \right],\nonumber\\
\end{eqnarray}

\begin{eqnarray}
 -\nabla   |\dot{\TTE}|^2 
  &=&
 -|\dot{\VEC{t}}|^2\nabla   |\xi|^2 
   =
 |\dot{\VEC{t}}|^2 \EE \frac{ \frac{d}{ds} |\xi|^2 } {|\dot{\VEC{r}}|},
\end{eqnarray}

\begin{eqnarray}
 -\nabla   \times\left( \xi|\dot{\VEC{t}}|\dot{\TRM}^* \right)
  &=&
-|\dot{\VEC{t}}| \left[ \nabla   \xi^* \times \dot{\TRM}^* + \xi^* \nabla   \times \dot{\TRM}^* \right]\nonumber\\
 &=&
 |\dot{\VEC{t}}| \left[ \EE \frac{\dot{\xi}^*}{|\dot{\VEC{r}}|} \times \dot{\TRM}^* + \xi^* \EM \times \frac{\ddot{\TRM}^*}{|\dot{\VEC{r}}|} \right]\nonumber\\
 &=&
 |\dot{\VEC{t}}| \left[ \EE \dot{\xi}^* \times \TEM^* + \xi^* \EM \times \dot{\TEM}^* \right].\nonumber\\
\end{eqnarray}

\subsection{From material science}
We should explain previous works in terms of complex coordinate method (CCM).
The complex coordinates are strongly studied in both mathematics and physics.\cite{ccm1,ccm2,ccm3}
Some of remarkable works successfully show the spectrum of excitated states in atoms by using the CCM.\cite{CCM1,CCM2,CCM3}
In the CCM, radial wave functions of solutions of the Schr\"odinger equation are solved along the axis deviating from real axis
  and hamiltonian matrix and eigenvalues become non hermitian and complex values, respectively.
From the deviations of the eigenvalues from real values,
  spectrum widths are additionally given as hidden information from the complex coordinates.
On the other hand, complex Gaussian basis having complex scale coefficients are also studied
  and used for some computational physics.\cite{CG1,CG2,CG3,CG4}
The most important thing we believe is that the product of two separated complex Gaussian basis
  are given by other complex Gaussian basis in other position in complex space.
So, the solutions of Schr\"odinger equation given by the complex Gaussian basis
  have those dispersions in complex space.

\end{document}